\documentclass[epj]{svjour}
\usepackage{graphicx}
\usepackage{amsmath}%

\usepackage{amsfonts}%
\usepackage{amssymb}%

\begin{document}
\institute{Physics Department, Brookhaven National Laboratory, Upton, NY, 11973,  USA}
\title{Quarkonium correlators and spectral functions at zero and finite 
temperature from Fermilab action}
\titlerunning{Quarkonium at zero and finite temperature}
\author{Konstantin Petrov}% etc
\authorrunning{K.Petrov}
\date{Received: date / Revised version: date}
\abstract{
We study charmonium and bottomonium systems at zero and finite temperatures using Lattice QCD with Fermilab action on anisotropic lattices.
 % end of PACS codes
} %end of abstract
\maketitle
\section{Introduction}
\label{intro}

There has been  considerable progress in studying quarkonium properties at finite 
temperature since the work of Matsui and Satz \cite{MS86}.
In the past the quarkonium properties at finite temperature were studied in potential models
\cite{karsch88,digal01a,digal01b} (for recent works see Refs. \cite{shuryak04,wong04}).
The applicability of the potential models at finite temperature is not obvious \cite{petreczkythis}.
It is more appropriate to study the in-medium modifications of meson properties at
finite temperature in terms of spectral functions (for recent review see \cite{rapp,petreczky03,Karsch:2003jg,petreczkylat04}).
Meson correlators in Eucledian time have been calculated in Lattice QCD for a long time, but 
to get spectral functions out of them  was considered impossible. 
However it was shown by
Asakawa, Hatsuda and Nakahara that using the \emph{Maximum Entropy Method} one
can in principle reconstruct also meson spectral functions. The method was
successfully applied at zero temperature \cite{asakawa01,cppacs} and later
also at finite temperature \cite{karsch02,wetzorke02,karsch03,asakawa04,datta04,asakawa02,umeda02,wetzorke03}. 
Though systematic uncertainties in the
spectral function calculated on lattice are not yet completely understood, it
was shown in Ref. \cite{karsch02,datta04} that a precise determination of the imaginary
time correlator can alone provide stringent constraints on the spectral
function at finite temperature.

\section{Lattice setup and simulation details}%

\label{sec:1}

We simulate both bottomonium and charmonium using the so called 
Fermilab action~\cite{Fermilab}. It is a special formulation of the O(a) improved Wilson action with 
broken axis-interchange symmetry: time-like and space-like
coefficients are treated independently and are determined in a non-perturbative
way using the dispersion relation. This formulation is known to significantly
reduce lattice artefacts at modest lattice spacings. As our major interest
lies at finite temperature theory we face the problem of time extent being
very short, which makes it difficult to fit the correlators. Therefore,
following \cite{chen01,Manke} we introduce anisotropic lattice, setting 
$a_s/a_t=\xi \neq 1$.

Such calculations require immense computational power, as to resolve ground
state from excited state we need small lattice spacing. One of the ways to
address this problem is building a dedicated Lattice QCD machine such as
QCDOC. This supercomputer was developed by physisists from Columbia
University, BNL, RIKEN and UKQCD. Three such machines, each reaching about
10TFlops peak performance, are currently under construction at BNL
and EPCC. We used for our simlations QCDOC prototypes, single-motherboard
machines at about 50 GFlops peak. Such resources are still not adequate for the
full QCD simulations, so we use quenched approximation, which is equivalent to
neglecting quark loops. Typical statistics gathered was 500 to 1000
measurements, separated by 400 updates.

To study meson properties at finite temperature one considers the following 
correlators of some operator $\hat{O}$
\begin{equation}
D^{>}(t,t^{\prime})=\langle\hat{O}(t)\hat{O}(t^{\prime})\rangle_{T}%
,~~~D^{<}(t,t^{\prime})=\langle\hat{O}(t^{\prime})\hat{O}(t)\rangle_{T},
\end{equation}
where $\langle...\rangle_{T}=\langle...e^{-\hat{H}/T}\rangle_{T}$ denotes the
thermal average \cite{lebellac}. As we cannot do any lattice simulation in
Minkowski space we need to introduce a Wick rotation and we get the imaginary time
correlator
\begin{equation}
G(\tau)=\langle\mathcal{T}\hat{O}(-i\tau)\hat{O}(0)\rangle.\label{rel1}%
\end{equation}
(where $\mathcal{T}$ denotes time ordering). Now we define a spectral
function through the Fourier transform of $D^{>(<)}(t)$ as
\begin{equation}
\displaystyle\sigma(\omega)=\frac{D^{>}(\omega)-D^{<}(\omega)}{2\pi}=\frac
{1}{\pi}ImD_{R}(\omega),
\end{equation}
where $D_{R}(\omega)$ is the retarded correlator. With the help of the KMS
condition on $D^{>(<)}$ \cite{lebellac} we arrive at the following integral
relation between the imaginary time correlator and the spectral function
\begin{eqnarray}
&
\displaystyle 
G(\tau)=\int_{0}^{\infty}d\omega\sigma(\omega)\frac{\cosh
(\omega(\tau-1/2T)}{\sinh\frac{\omega}{2T}}\nonumber\\
&
\displaystyle 
\equiv\int_{0}^{\infty}
d\omega\sigma(\omega)K(\omega,\tau)
\label{rel2}
\end{eqnarray}
Now we can try to reconstruct the spectral function by calculating $G(\tau)$
on the lattice; however a set of complications arises: At zero temperature
the kernel $K(\omega,\tau)$ reduces to simple exponential and when we consider
large Euclidean times we only see the contribution from the lowest lying meson
state in $\sigma(\omega)$, i.e $G(\tau)=\exp(-m\tau)$. However at finite
temperature the time interval is limited and excited states are as important
as the ground state. Additional problems arise in lattice calculations where
correlators are calculated only on a discrete set of Euclidean times $\tau
T=k/N_{\tau}$, $k=0,...,N_{\tau}-1$ with $N_{\tau}$ being the temporal extent
of the lattice. In order to reconstruct the spectral functions from this
limited information it is necessary to include in the statistical analysis of
the numerical results also prior information on the structure of
$\sigma(\omega)$ (e.g. such as $\sigma(\omega)>0$ for $\omega>0$). This can be
done in many ways, none of them being rigorous. Therefore we use two of such
methods, namely the \emph{Maximum Entropy Method} (MEM) \cite{nakahara99,asakawa01}
and constrained curve fitting \cite{lepage}\footnote{Other
methods of introducing prior information into the statistical analysis has also 
been  discussed in Ref. \cite{gupta-v}.}.
These two methods are totally different both in their assumptions and
the procedures and we use them to cross-check our results.

To do a lattice study of particles with given quantum numbers
we must produce an
appropriate set of operators with given symmetry properties. One such set is a
local meson operator which is bilinear in quark-antiquark fields (current)
\cite{asakawa01,cppacs,karsch02,wetzorke02}
\begin{equation}
O_{H}(\tau,\vec{x})=\bar{q}(\tau,\vec{x})\Gamma_{H}q(\tau,\vec
{x}),~~~\Gamma_{H}=1,\gamma_{5},\gamma_{\mu},\gamma_{5}\gamma_{\mu
}\label{pointop}
\end{equation}
for scalar, pseudoscalar, vector and axial vector channels correspondingly.
The pseudo-scalar and vector correlators correspond to ground state 
quarkonia $\eta_{c,b}(^1 S_0)$ and \\
$J/\psi,\Upsilon(^3 S_1)$ respectively.
The scalar and axial-vector channels correspond to P state quarkonia, $\chi_{c,b}(J=0,1)$.
The temporal
correlators at finite spatial momentum $\vec{p}$ then take the following form%
\begin{eqnarray}
&
G_{H}(\tau,\vec{p})=\langle O_{H}(\tau,\vec{p})O_{H}^{\dagger}(\tau,-\vec
{p})\rangle, \nonumber\\
&
O(\tau,\vec{p})=\sum_{\vec{x}}e^{i\vec{p}\vec{x}}O_{H}(\tau,\vec{x})
\end{eqnarray}

\section{Charmonium at zero and finite temperature}

\begin{figure}
\includegraphics[width=3.5in]{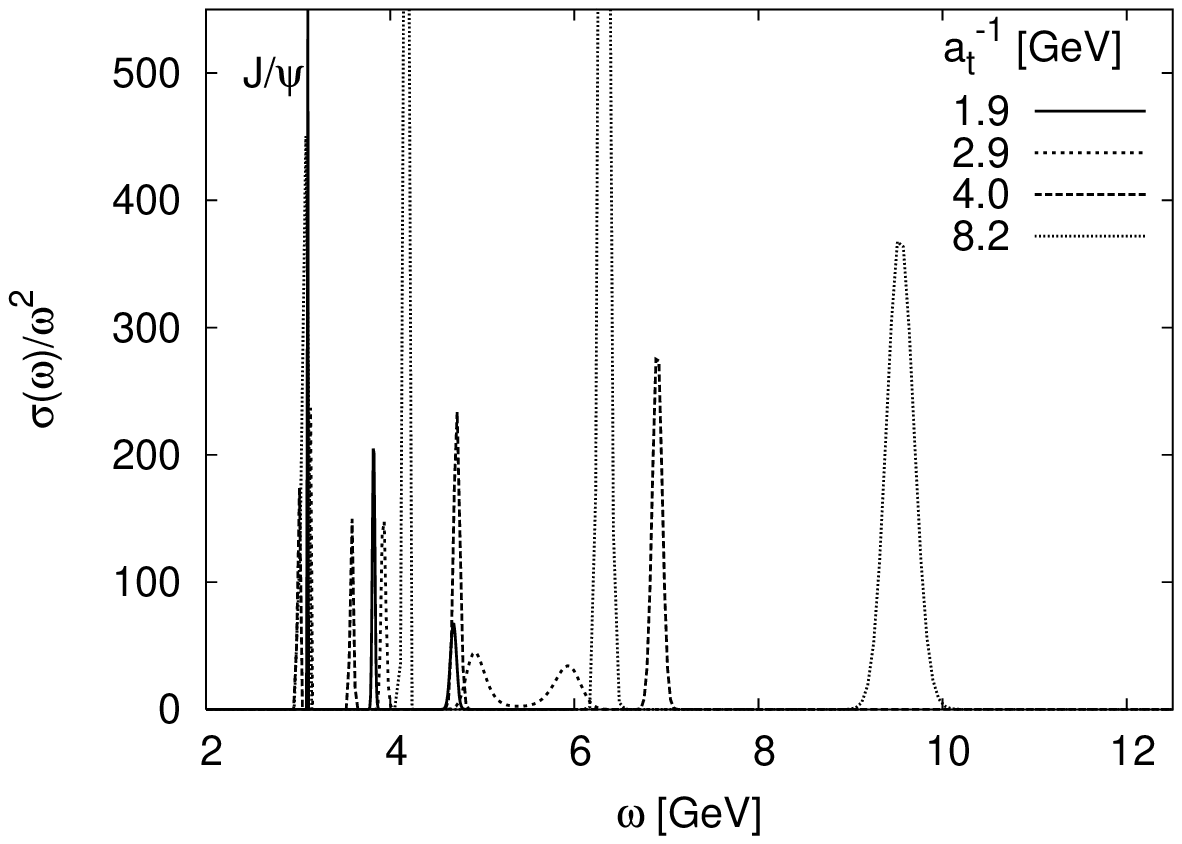}
\includegraphics[width=3.5in]{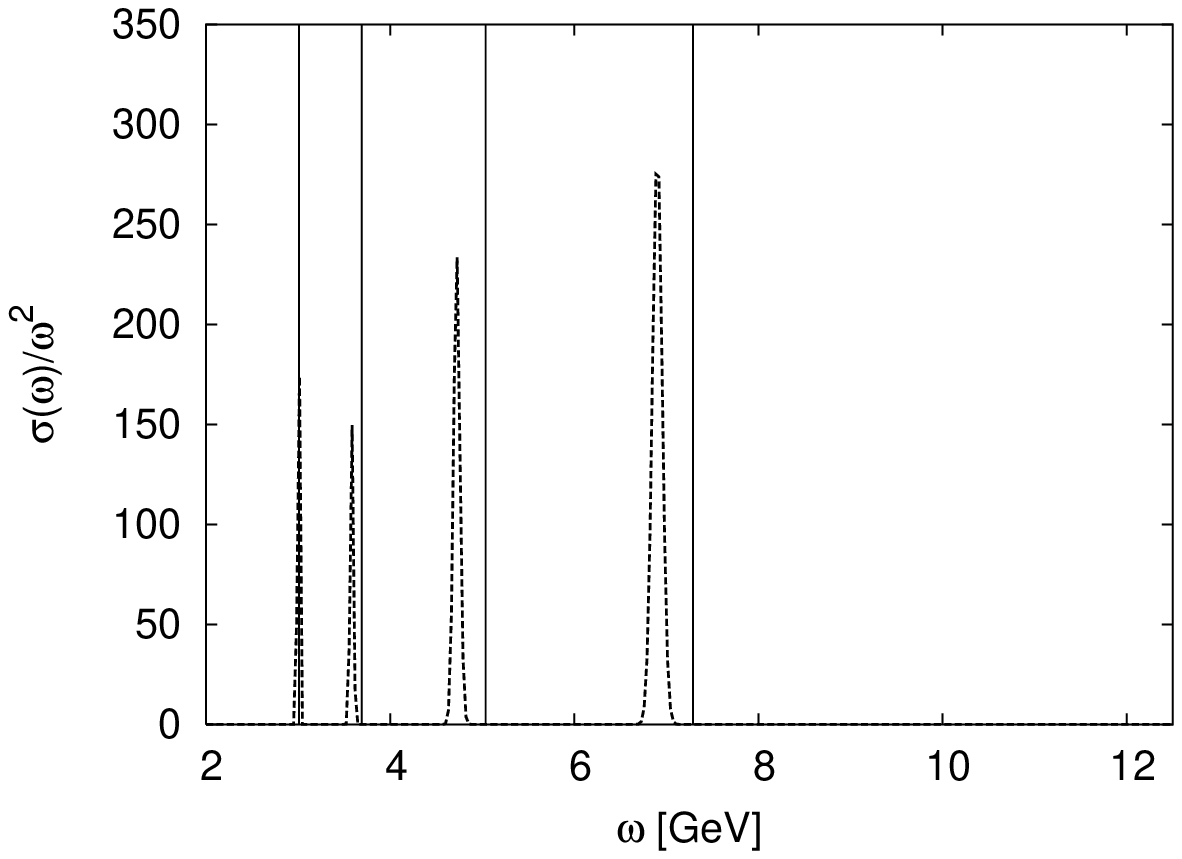}
\label{c_zero}
\caption{The zero temperature 
vector spectral function for charmonium at different lattice spacings (top).
Also shown the vector spectral function at $a_t^{-1}=4$GeV compared with the results
of constrained curve fit (vertical lines). }
\end{figure}

We start the discussion of the numerical results with zero temperature 
spectral functions in the vector channel.  In Fig. \ref{c_zero} 
we show the spectral functions at four different lattice spacings
$a_t^{-1}=1.9,~2.9,~4.0$ and $8.2$ GeV. The anisotropy $\xi=2$ for the 
first three lattice spacing and $\xi=4$ for the last one. The lattice spacing
was fixed using the heavy quark potential and the value of $r_0=0.5$ fm for
the Sommer scale (see \cite{chen01} for further details). As one can
see from the figure the first peak, which corresponds to the $J/\psi$ state, 
does not move as we vary the lattice spacings. 
On the contrary, the position of the
other peaks depends on the lattice spacing. Moreover, as we go to finer
lattices more peaks appear. 
Similar results were obained also in other channels. Thus, all the structures except the first peak
cannot be identified with physical states. It is possible that these peaks
in fact belong to the continuum distorted by effects of finite lattice. In any case 
this problem requires further analysis which will be presented elsewhere \cite{self}.
We also analyzed the spectral functions using constrained curve fitting
by using a multi-exponential Ansatz with four or more terms.
The results from the fits are shown in Fig.\ref{c_zero}.
For the ground state we get very good agreement between MEM and constrained fit, 
while it becomes worse for higher states. Nonetheless, the rough agreement between
the two methods gives us some confidence as well as indicating the size of possible systematic
errors.  

Now we are in a position to discuss spectral functions at finite temperature. 
For $a_t^{-1}=8.2$ GeV we have sufficient number of data points for temperatures
above deconfinement to analyze the spectral function.  
In Fig.2
the spectral function in pseudo-scalar and scalar channel is shown.  The 
peak corresponding to $\eta_c$ state survives in the plasma phase; moreover, its
position is essentially unchanged. This is consistent with previous findings \cite{asakawa04,datta04,umeda02}. 
The scalar channel has been analyzed only in Ref.\cite{datta04} and it was found that the $\chi_c$ state
dissolves in the plasma soon after the deconfinement temperature.
The analysis of the scalar spectral function shown in Fig.2 confirms this
conclusions.
\begin{figure}
\includegraphics[width=2.5in, angle=-90]{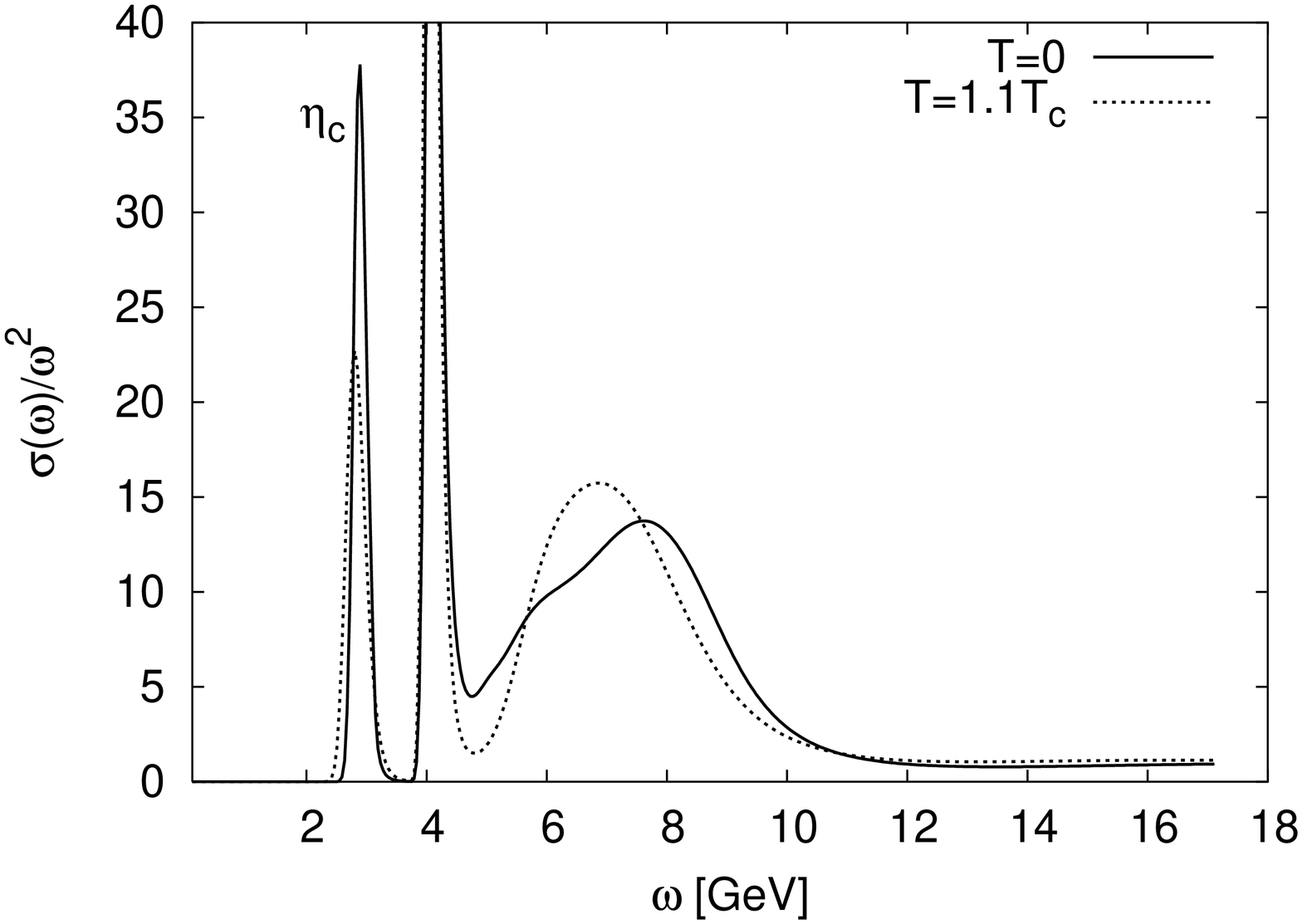}
\includegraphics[width=2.5in, angle=-90]{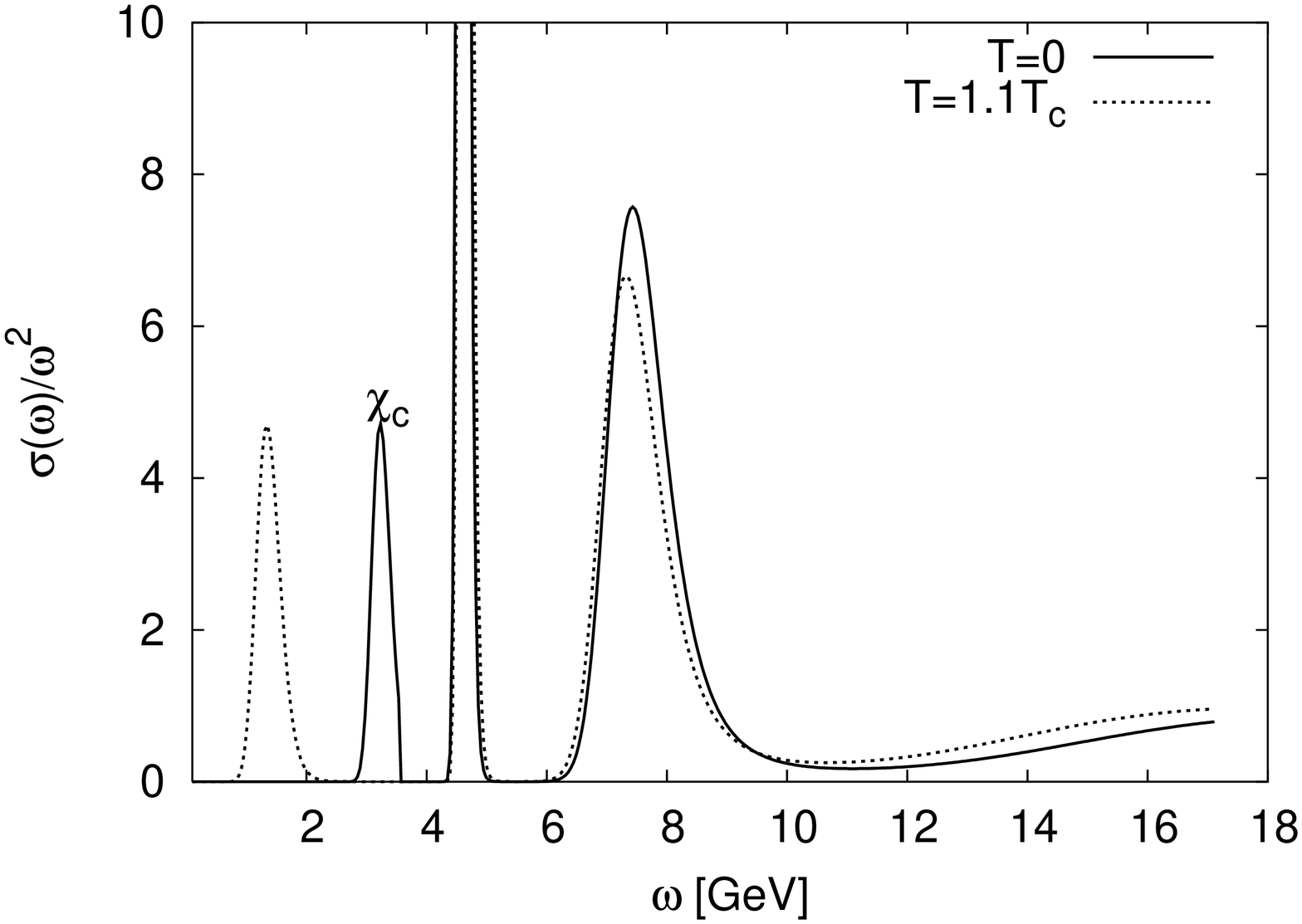}
\label{c_finite}
\caption{The charmonia spectral functions for $a_t^{-1}=8.2$ GeV at zero temperature and above deconfinement
temperature $T=1.1T_c$ corresponding to $N_t=24$. }
\end{figure}

\section{Bottomonium spectral functions and correlators}

\bigskip While charmonium has been studied by other collaborations, the bound
states of $b$-quarks received much less attention. In fact this is the first
study of the bottomonium at finite temperature. As in the charmonium case we
start with zero temperature physics. 
The bottominium spectrum using Fermilab action was studied in Ref. \cite{Manke}
where extened meson operators have been used. We have studied the bottomonium spectrum
using two lattice spacings $a_t^{-1}=8.2$GeV and $a_t^{-1}=12.0$GeV and $\xi=4$ estimated from
the Sommer scale\footnote{In Ref. \cite{Manke} the 1P-1S splitting was used to determine
the lattice spacing which leads to a largely overestimated value for it.}.
In Fig.3  bottomonia spectral functions in scalar and pseudo-scalar channels
are shown. Because of the large bottomonia mass the correlators are quite noisy and 
we found that it is considerably more difficuilt to reconstruct the spectral functions then it was in
the charmonia case. 
\begin{figure}
\includegraphics[width=3.5in]{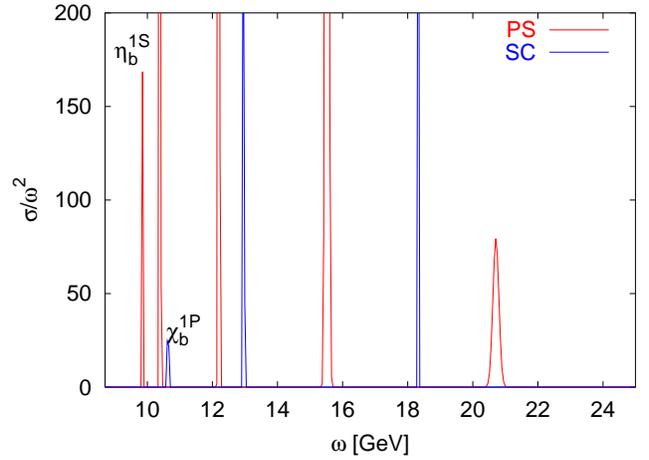}
\label{fig:botzero}
\caption{Bottomonium spectral functions at zero temperature, scalar and pseudo-scalar channels
for $a_t^{-1}=12$GeV.}
\end{figure}

\begin{figure}
\includegraphics[width=2.9in]{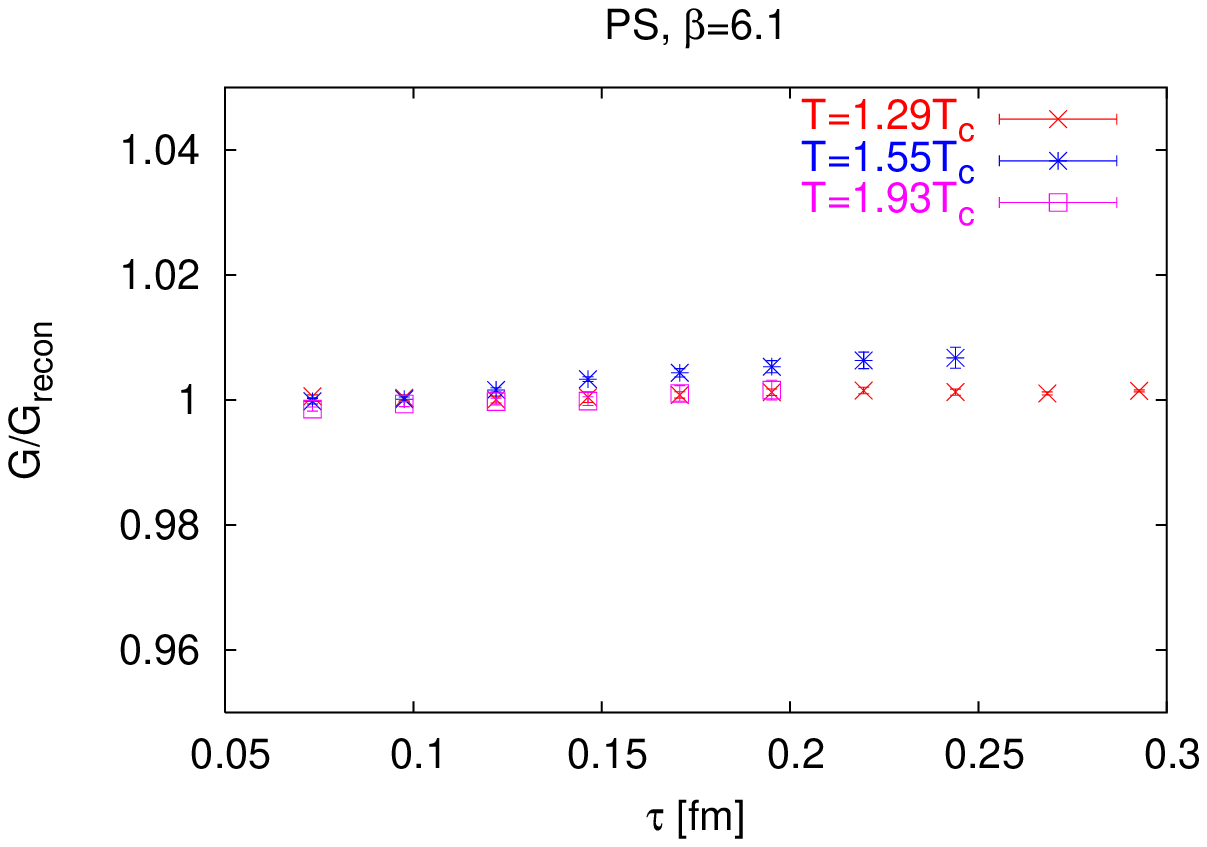}
\includegraphics[width=2.9in]{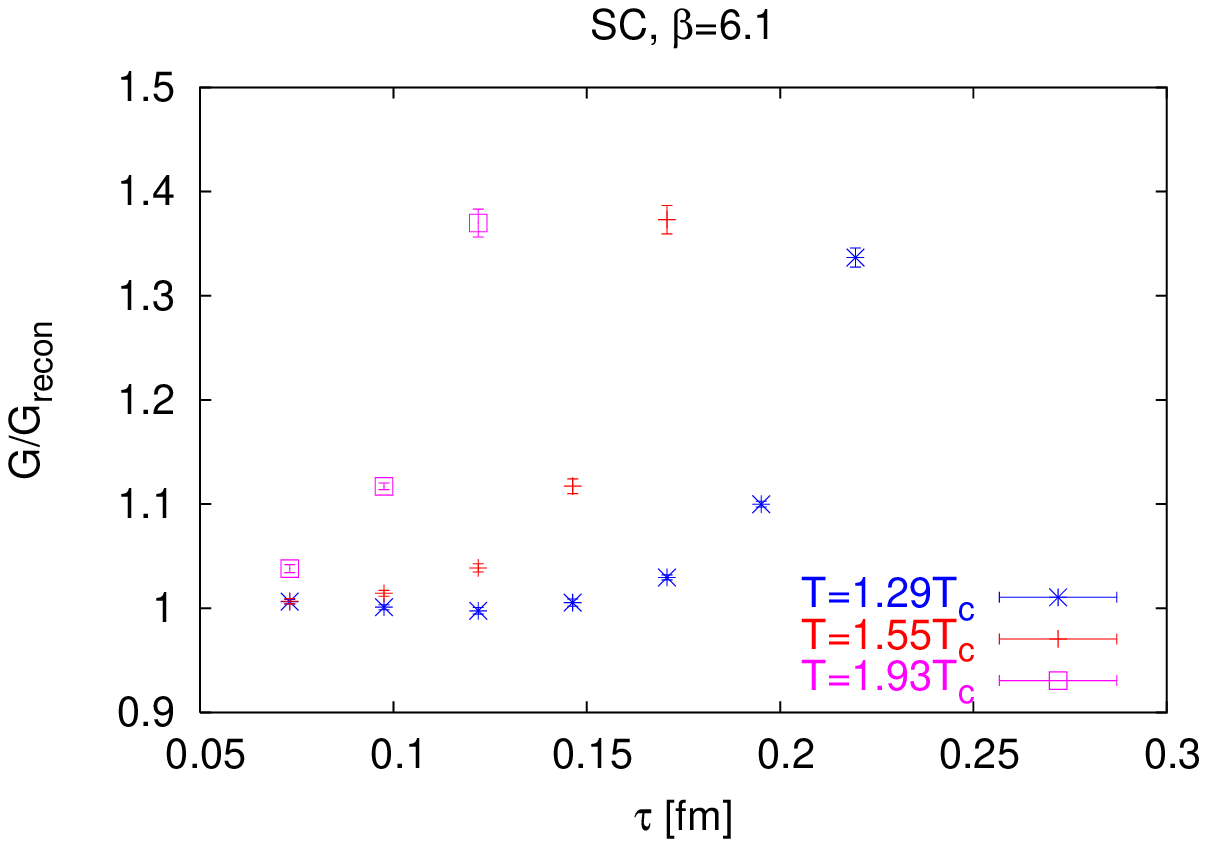}
\label{fig:bot:corr}
\caption{
The ratio $G(\tau,T)/G_{recon}(\tau,T)$ at different temperatures for 
pseudo-scalar (top) and scalar (bottom) correlators.
}
\end{figure}

As the analysis of the bottomonium spectral functions is difficuilt already at zero
temperature at finite temperature we study only the temperature 
dependence of the correlator. 
This method is based on the following argument \cite{datta04}: in Eq.\ref{rel2} the
temperature dependence of the right hand side comes from two sources - the
spectral function itself and the finite temperature kernel. Now let us
introduce the so-called reconstructed correlator using the spectral function at
zero temperature
\begin{equation}
\displaystyle G_{recon}(\tau,T)=\int_{0}^{\infty}d\omega\sigma(\omega
,T=0)\frac{\cosh(\omega(\tau-1/(2T))}{\sinh\frac{\omega}{2T}}.
\end{equation}
If the spectral function does not depend on temperature then the reconstructed
and directly calculated correlators will be equal, i.e. $G(\tau,T)/G_{recon}(\tau,T)=1$.
We show their ratio for the pseudo-scalar channel ($S-wave$) and
the scalar channel ($P-wave$)  of the bottomonium on Fig.\ref{fig:bot:corr}. 
As one can see there is no modification
of the correlator for the $\eta_{b}$ particle till almost twice the critical
temperature. However, quite suprisingly, for the $\chi_{b}$ we see
a drastic change already at $1.3T_{c}$ which suggests that something has
happened. The $\chi_b$ state has approximately the same size and binding energy
as the $J/\psi$ or $\eta_c$ state which do not show any change till $(1.7-2.0)T_c$.
Thus we would expect the same for $\chi_b$. It  also is possible that the large change in the scalar correlator is not caused by 
the dissolution of the  $\chi_b$ but rather by modification of the continuum part of the spectral function \cite{mocsy}.

\section{Conclusions and Outlook}

We analysed charmonium and bottomonium systems at zero and finite temperatures
using Lattice QCD with Fermilab action on anisotropic lattices. We
reconstructed their spectral functions, reliably identifying ground states for
various values of the lattice cutoff. The $S-wave$ states in both systems survive in
the deconfined phase till as much as three times the critical temperature,
while $P-wave$ states may dissolve shortly after the phase transition. The situation with
higher excited states, especially $2S$, is very questionable and requires further study.

\section{Acknowlegements} This presentation is done based on work in collaboration with P.~Petreczky and A.~Velytsky 
and is supported by U.S. Department of Energy under Contract No. DE-AC02-98CH10886 and by SciDAC project. 
The Maximum entropy method analysis was done using the program developed by A. Jakov\'ac.
K.P. would like to thank Kavli ITP at
UCSB for the hospitality. Results are obtained on RIKEN/BNL QCDOC prototypes
using Columbia Physics System with high-performance clover inverter by P.~Boyle
and other parts by RBC collaboration. Special thanks to C.~Jung for his generous help with CPS.

\end{document}